# Influencing factors of Twitter mentions of scientific papers

Pablo Dorta-González

University of Las Palmas de Gran Canaria, TiDES Research Institute, Campus de Tafira, 35017 Las Palmas de Gran Canaria, Spain. E-mail: pablo.dorta@ulpgc.es ORCID: http://orcid.org/0000-0003-0494-2903

**Abstract**

**Purpose:** This paper explores some influencing factors of Twitter mentions of scientific research. The results can help to understand the relationships between various altmetrics.

**Design/methodology/approach:** Data on research mentions in Altmetric and a multiple linear regression analysis are used.

**Findings:** Among the variables analyzed, the number of mainstream news is the factor that most influences the number of mentions on Twitter, followed by the fact of dealing with a highly topical issue such as COVID-19. The influence is weaker in the case of expert recommendations and the consolidation of knowledge in the form of a review. The lowest influence corresponds to both the public policies through references in reports, and to citations in Wikipedia, while mentions in patent applications does not have a significant influence.

**Research limitations:** A specific field was studied in a specific time frame. Studying other fields and/or different time periods might result in different findings.

**Practical implications:** Governments increasingly push researchers toward activities with societal impact and this study can help understand how different factors affect social media attention.

**Originality/value:** Understanding social media attention of research is essential when implementing societal impact indicators.

## 1. Introduction

Governments increasingly push researchers toward activities with societal impact, including economics, cultural and health benefits (Thelwall, 2021). That is why since the term 'altmetrics' was introduced in 2010 (Priem et al., 2010), theoretical and practical research have been conducted in this discipline (Sugimoto et al., 2017). Thus, altmetrics allow knowing how the results of scientific research are perceived and commented beyond academia (Dorta-González and Dorta-González, 2022).

Twitter is nowadays the most used social media platform by the general population (and by researchers in particular) to disseminate and comment on the results of scientific research. There are many and varied factors that may influence social media attention of research. These factors include the mainstream news coverage, the topic addressed (COVID-19, for example), and some characteristics of the research such as its proximity to social issues (impact on public policy) and business (impact on patents), their contribution to the consolidation of knowledge (in the form of review or mention on Wikipedia), and the recommendation of experts (Faculty Opinions).

In this paper, the effect of these factors on the social attention that research receives through mentions on Twitter is quantified. For this, data on mentions in Altmetric and a multiple linear regression analysis are used. The unit of study is the research paper (article and review) in disciplinary journals in the field of Clinical Medicine. The time frame covers 2018-2020 and the country analyzed is Spain.

## 2. Theoretical framework on altmetrics

Most altmetric data improve citations regarding the accumulation speed after publication (Fang and Costas, 2020). However, except for Mendeley readership which is moderately correlated with citations (Zahedi and Haustein, 2018), there is a negligible or weak correlation between citations and most altmetric indicators (Bornmann, 2015; Costas et al., 2015). This means that altmetrics might capture diverse forms of impact which are different from citation impact (Wouters et al., 2019).

Altmetrics come to cover the need for researchers to provide evidence of the societal impact of their results. However, it is difficult to measure the societal impact of research because a long time can elapse between basic research and its practical applications (Godin, 2011), and because the obsolescence of the results strongly determines the impact of research (Dorta-González and Gómez-Déniz, 2022). Thus, in addition to mentions in social media and mainstream news, altmetrics also include references in public policy documents and recommendations more scholarly than societal (Haustein et al., 2016). Furthermore, the variety of indicators and their differences advise using them separately instead of mixed indicators (Wouters et al., 2019).

The attention received by research and its impact are not synonymous. Social attention is a more complex phenomenon because it can be motivated by positive or negative aspects of the research (Sugimoto et al., 2017). Among the different dimensions of social attention, the following can be mentioned. Mentions on Twitter and Facebook can represent discussion on social media, blogs and news might reflect attention about newsworthiness, and Wikipedia might explain informational attention (Thelwall and Nevill, 2018). Moreover, there are different levels of attention depending on the commitment that the social interaction entails. In this way, it is not the same to retweet than to write a post on a blog (Haustein et al., 2016).

Authors found a higher presence of altmetrics in social sciences and humanities than in the natural sciences (Chen et al., 2015). Thus, they suggested that altmetrics can represent a complement to citations, especially in humanities and social sciences.

Authors have also been interested in the platforms that collect altmetrics, such as Altmetric, Impactstory, and PlumX, in relation to the data source, the indicators provided, the speed of data accumulation, and so on (Fang and Costas, 2020; Fang et al., 2020; Ortega, 2018). The differences between countries and disciplines according to the coverage of the mentions have also been analyzed (Torres-Salinas et al., 2022).

As indicated, there are many and varied factors that influence social attention of research. In addition to those already mentioned, Faculty Opinions (formerly F1000Prime) is a system for post-publication peer-review, in which experts identify, assess, and comment relevant papers they read (Bornmann and Haunschild, 2018).

## 3. Materials and Methods

Disciplinary journals correspond to those in the field of Clinical Medicine in Web of Science database. To identify the scientific production of Spain, all research articles and reviews in Web of Science which at least one co-author had affiliation located in Spain were considered. For it, the search field 'Address' was used (AD=Spain), and the query was limited to the Science Citation Index Expanded (SCI-Expanded).

Web of Science was also used to identify those papers that include the term COVID-19 and/or SARS-CoV-2 in the title, and the document typology.

The source of altmetrics data is Altmetric. This is a currently popular and one of the first altmetrics aggregator platforms, that originated in 2011 with the support of Digital Science. It tracks and accumulates mentions from different social media platforms, news, blogs, and other sources for mentions of scholarly articles (Altmetric, 2020).

The altmetrics data were identified through the DOI of the document. For this, the DOI of each publication in the sample was searched in Altmetric. The altmetric data measures the social attention a paper receives from mainstream and social media, public policy documents, Wikipedia, etc. It collects the online presence and analyzes the conversations around the research.

Thus, the Web of Science database provided the DOIs of the Spanish research production in Clinical Medicine in 2018-2020, a total of 47,211 research papers of which 29,894 were in the Altmetric platform with some social mention (60%). A total of 610 papers of those indexed in Altmetric contained the terms COVID-19 and/or SARS-CoV-2 in the title. Data were retrieved in December 2021.

It was decided to use a random sample because the number of records exceeded the calculation capacity. To do this, a set of numbers was randomly generated and associated with the records in the dataset.

The sample employed in this study is described in Table 1. I decided to add the COVID-19 papers to a simple random sample of the other papers. In this way, the years 2018 and 2019 correspond to a simple random sample, while the year 2020 is a compendium of a random sample and the COVID-19 papers. The COVID-19 papers are therefore overrepresented in the sample, 12.5% versus 2% in the population. This decision is motivated by the fact that, except for news and mentions on Twitter, the rest of the variables analyzed are very infrequent (see Table 3) and, therefore, taking only 2% of COVID-19 publications means that some variables are mostly equal to zero within the COVID-19 group and, therefore, would not explain anything in the proposed model. Thus, the resulting sample size was N= 4895.

Table 1

*Random sample description.*

|  | Papers | Group | Year | | | | |
|---|---|---|---|---|---|---|---|
|  |  |  | 2018 | 2019 | 2020 | 2018-2020 | % (of Total) |
| Altmetric.com | 29,894 | COVID-19[1] | 0 | 0 | 610 | 610 | 2.0% |
|  |  | Others | 9064 | 9894 | 10,326 | 29,284 | 98.0% |
|  |  | Total | 9064 | 9894 | 10,936 | 29,894 |  |
| Sample | N=4895 | COVID-19[1] | 0 | 0 | 610 | 610 | 12.5% |
|  |  | Others | 1305 | 1506 | 1474 | 4285 | 87.5% |
|  |  | Total | 1305 | 1506 | 2084 | 4895 |  |
| % (of Population) | 16.4% |  | 14.4% | 15.2% | 19.1% |  |  |

[1] COVID-19 or SARS-CoV-2 in the title

The methodology in this paper consists of a Multiple Linear Regression analysis. So, the dependent variable is the social media attention of a research measured through the number of tweets and retweets in Twitter, or Twitter mentions for short. The independent variables are described in Table 2.

Regarding the selection of independent variables, the following considerations can be made. Other possible altmetric indicators and bibliometric variables could be consider. The criterion when selecting the independent variables of the model was to incorporate different dimensions. Thus, mentions on social media measure the same dimension. Mentions on Facebook are much less frequent than on Twitter, and, unlike other studies, I chose to use a single indicator instead of an aggregation of measures with arbitrary weights.

Table 2

*Variables and description in the Multiple Linear Regression model.*

|   | Name | Variable | Description | Source | Type |
|---|------|----------|-------------|--------|------|
| 1. | Twitter | Social media attention | Number of mentions in the social media Twitter (tweets and retweets including the DOI of the paper) | Altmetric.com | Natural number N = {0, 1, 2, …} |
| 2. | News | Mainstream news | Number of news in the mainstream media | Altmetric.com | Natural number N = {0, 1, 2, …} |
| 3. | Patent | Patent application mentions | Number of mentions in patent applications | Altmetric.com | Natural number N = {0, 1, 2, …} |
| 4. | Policy | Public policy mentions | Number of mentions in public policy documents | Altmetric.com | Natural number N = {0, 1, 2, …} |
| 5. | Wikipedia | Wikipedia mentions | Number of mentions in Wikipedia articles | Altmetric.com | Natural number N = {0, 1, 2, …} |
| 6. | COVID-19 | COVID-19 in the title | Inclusion of COVID-19 or SARS-CoV-2 in the paper title | Web of Science | Dichotomous {yes = 1, not = 0} |
| 7. | F1000 | Expert mentions | Number of recommendations on Faculty Opinions (formerly F1000Prime) | Altmetric.com | Natural number N = {0, 1, 2, …} |
| 8. | Review[1] | Type of research | Typology of the paper | Web of Science | Dichotomous {review = 1, article = 0} |

[1]A systematical review of the literature puts research into context for news reporters, policymakers, the public and other researchers.

Something similar happens with blogs and news. Both correspond to the same dimension. The correlation between both variables is high, so it was decided to include only the news instead of an aggregation with arbitrary weights.

Regarding the bibliometric indicators, a current case of study (Covid-19) well represented in the field analyzed (Clinical Medicine) was used. Although open access might be of interest, during the pandemic publishers gave open access to all publications on Covid-19, so it was not finally included in the model.

About the review documentary typology, keep in mind that a systematical review of the literature puts research into context for news reporters, policymakers, the public, and other researchers. In this sense, it seems relevant a priori.

Finally, scientific collaboration is an interesting aspect because it increases both citations and mentions. However, this aspect would require a more detailed study in relation to the type of collaboration (number of co-authors, number of affiliations, number of countries, etc.)

## 4. Results

The objective of the study is to know if and how social media attention of a scientific research, taking as proxy the number of mentions in Twitter, can be explained from other social mentions and bibliometric characteristics of the paper.

So, the dependent variable is the number of mentions in Twitter. The independent variables are the following: mainstream news; references in patent applications, public policy documents, and Wikipedia articles; expert mentions (taking as proxy the number of recommendations in Faculty Opinions -formerly F1000Prime-); the topic through the inclusion of COVID-19 or SARS-CoV-2 in the title; and the publication typology (article and review). The description of the variables is shown in Table 2.

I checked the distributions in the histograms are reasonable for all variables (see Table 3 for the mean and standard deviation, among other descriptive measures). Note there are N = 4895 independent observations in our dataset. I checked also for curvilinear relations or anything unusual in the plot of the dependent variable versus each independent variable.

Table 3

*Descriptive of the random sample.*

| Variable | Mean | Median | Mode | SD | Minimum | Maximum | Sum | Count |
| --- | --- | --- | --- | --- | --- | --- | --- | --- |
| 1. Twitter | 47.117 | 6 | 1 | 463.129 | 0 | 15,695 | 230,637 | 4895 |
| 2. News | 2.298 | 0 | 0 | 27.489 | 0 | 1429 | 11,247 | 4895 |
| 3. Patent | 0.016 | 0 | 0 | 0.186 | 0 | 6 | 80 | 4895 |
| 4. Policy | 0.040 | 0 | 0 | 0.414 | 0 | 20 | 194 | 4895 |
| 5. Wikipedia | 0.025 | 0 | 0 | 0.226 | 0 | 6 | 120 | 4895 |
| 6. COVID-19[1] | 0.125 | 0 | 0 | 0.330 | 0 | 1 | 610 | 4895 |
| 7. F1000 | 0.021 | 0 | 0 | 0.174 | 0 | 4 | 103 | 4895 |
| 8. Review[2] | 0.146 | 0 | 0 | 0.353 | 0 | 1 | 716 | 4895 |

Note: [1] COVID-19 or SARS-CoV-2 in the title (yes = 1, not = 0). [2] Type (review = 1, article = 0)

Since the considered variables have skewed distributions (see Table 3), the Spearman correlations are shown in Table 4. Note the independent variables have a statistically significant relation with Twitter mentions, but small in many cases (see Table 4). Twitter mentions correlate mainly with news (0.30) and COVID-19 (0.19). To a lesser extent they also correlate with F1000 (0.14), policy (0.11), and Wikipedia (0.10).

Therefore, the multiple linear regression model could estimate the Twitter mentions from all independent variables simultaneously. For it, I checked the correlations among the independent variables (Table 4). Note all the absolute correlations are low (none of them exceed 0.19) and multicollinearity problems are discarded for the actual regression analysis.

Note the only significant negative correlations are observed between COVID-19 and patent (-0.04), and between review and F1000 (-0.03). Although very low, they are statistically different from zero. This means that articles on the COVID-19 topic are less referenced in patent applications and that review articles receive fewer expert recommendations. In the first case, it may be because all the COVID-19 articles correspond to the last year of the period analyzed and have had less time to be incorporated into patent applications.

Table 4

*Spearman correlations among the dependent and all independent variables.*

|    |           | 1 | 2     | 3     | 4     | 5     | 6       | 7     | 8     |
|----|-----------|---|-------|-------|-------|-------|---------|-------|-------|
| 1. | Twitter   | - | .30** | .03*  | .11** | .10** | 0.19**  | .14** | .03*  |
| 2. | News      |   | -     | .11** | .15** | .15** | .14**   | .19** | -.03  |
| 3. | Patent    |   |       | -     | .05** | .06** | -.04**  | .12** | .00   |
| 4. | Policy    |   |       |       | -     | .13** | .13**   | .13** | .00   |
| 5. | Wikipedia |   |       |       |       | -     | .04**   | .12** | .05** |
| 6. | COVID-19[1] |   |     |       |       |       | -       | .01   | .02   |
| 7. | F1000     |   |       |       |       |       |         | -     | -.03* |
| 8. | Review[2] |   |       |       |       |       |         |       | -     |

Note: *$p < 0.05$. **$p < 0.01$. [1] COVID-19 or SARS-CoV-2 in the title (yes = 1, not = 0). [2] Type (review = 1, article = 0)

According to the b-coefficients in Table 5, the regression model is:

$$\text{Log}_{10}(\text{Twitter}_i) = 0.721 + 0.824 \cdot \log_{10}(\text{News}_i) - 0.203 \cdot \log_{10}(\text{Patent}_i) \quad (1)$$
$$+ 0.534 \cdot \log_{10}(\text{Policy}_i) + 0.747 \cdot \log_{10}(\text{Wikipedia}_i)$$
$$+ 0.332 \cdot \text{COVID-19}_i + 0.235 \cdot \text{F1000}_i + 0.099 \cdot \text{Review}_i$$

where $\text{Twitter}_i$ denotes predicted Twitter mentions for paper i, i=1,2, …, 4895.

The adjusted R-square is reported in Table 5. R-square is the proportion of variance in the dependent variable accounted by the model. In our model $R^2_{adj} = 0.213$, which is considered acceptable in social sciences for a model that does not pretend to predict but to explain a social phenomenon. Moreover, since the p-value found in the ANOVA is less than $10^{-4}$, the entire regression model has a non-zero correlation.

Table 5

*Explanatory Model. Regression coefficients for predicting Twitter mentions. Standard Multiple Linear Regression analysis.*

| Variable | B (Coeff.) | 95% CI | β (Standardized Coeff.) | t | p (Sig.) |
|---|---|---|---|---|---|
| Constant | 0.721 | [0.701, 0.741] | 0.000 | 71.793 | 0.000 |
| $Log_{10}$(News) | 0.824 | [0.761, 0.887] | 0.366 | 25.636 | 0.000 |
| $Log_{10}$(Patent) | -0.203 | [-0.917, 0.510] | -0.007 | -0.559 | 0.576 |
| $Log_{10}$(Policy) | 0.534 | [0.111, 0.958] | 0.034 | 2.473 | 0.013 |
| $Log_{10}$(Wikipedia) | 0.747 | [0.141, 1.353] | 0.033 | 2.418 | 0.016 |
| COVID-19[1] | 0.332 | [0.280, 0.384] | 0.162 | 12.604 | 0.000 |
| F1000 | 0.235 | [0.131, 0.339] | 0.061 | 4.435 | 0.000 |
| Review[2] | 0.099 | [0.051, 0.146] | 0.051 | 4.051 | 0.000 |

Note. Adjusted R-square $R^2_{adj}$ = 0.213 (N = 4895, $p < 10^{-4}$). CI = confidence interval for B. [1] COVID-19 or SARS-CoV-2 in the title (yes = 1, not = 0). [2] Type (review = 1, article = 0)

Note each b-coefficient in equation (1) indicates the average increase in Twitter mentions (in a base-10 logarithmic scale) associated with an increase of ten units in those predictors in logarithmic scale or associated with a unit increment in the other predictors, everything else equal.

Thus, ten additional mainstream news are associated with a 6.67 (potency in base ten of 0.824) average increase in the number of mentions in Twitter, everything else equal. Similarly, ten additional policy mentions increase Twitter mentions on average by 3.42 (i.e., $10^{0.534}$). Moreover, ten additional Wikipedia mentions are associated with a 5.59 (i.e., $10^{0.747}$) average increase in the number of mentions in Twitter, ceteris paribus. Note ten additional patent mentions contribute an average 1.60 (i.e., $10^{0.203}$) decrease in Twitter mentions. However, this coefficient was not statistically significant. Analogously, each expert mention F1000 is associated with a 1.72 (i.e., $10^{0.235}$) average increase in the number of mentions in Twitter (72% higher for each recommendation), everything else equal.

About the dichotomous variables, a 1-unit increase in COVID-19 results in an average 2.15 (i.e., $10^{0.332}$) mentions increase in Twitter. Note that COVID-19 is coded 0 (not) and 1 (yes) in our dataset. So, for this variable, the only possible 1-unit increase is from not COVID-19 to COVID-19. Therefore, the average Twitter mentions for COVID-19 papers is 2.15 times higher than for not COVID-19 papers (more than double), everything else equal. Similarly, the average Twitter mentions for a review paper is 1.25 (i.e., $10^{0.099}$) times higher than for research papers (25% higher), ceteris paribus.

The statistical significance column (Sig. in Table 5) contains the 2-tailed p-value for each b-coefficient. Note that most b-coefficients in the model are highly statistically significant with a p-value less than $10^{-4}$. However, mentions in patent applications does not have a significant influence.

Note the b-coefficients don't indicate the relative strengths of predictors. This is because independent variables have different scales. The standardized regression coefficients or beta coefficients, denoted as β in Table 5, are obtained by standardizing all regression variables before computing the coefficients, and therefore they are comparable within and between regression models.

Thus, the two strongest predictors in the coefficients are the news (β = 0.366) and the topic COVID-19 (β = 0.162). This means that the number of news is the factor, among those analyzed, that contributes the most to the social media attention of a research (Twitter mentions), approximately 2.3 times more than a very media topic as COVID-19. Moreover, the number of news contributes to Twitter mentions 6 times more than expert mentions F1000 (β = 0.061), 7.2 times more than the review typology (β = 0.051), 10.8 times more than mentions in policy documents (β = 0.034), and 11.1 times more than mentions in Wikipedia (β = 0.033).

About the multiple regression assumptions, each observation corresponds to a different paper. Thus, I consider them as independent observations. The regression residuals are approximately normally distributed in the histogram. I also checked the homoscedasticity and the linearity assumptions in a plot of residuals versus predicted values. This scatterplot does not show any systematic pattern and therefore both assumptions hold.

Therefore, as main conclusion, among the dimensions analyzed in this paper, the factor that contributes the most to the attention in social media (Twitter mentions) is the number of news, followed by the topic (COVID-19), 44% in relation to the news when comparing documents with similar characteristics. Other factors that also positively affect social attention in Twitter, although on a smaller scale, are expert mentions F1000 (17% compared to news), the review typology (14% compared to news), and mentions in policy documents and Wikipedia (9% compared to news). All these relationships in average terms and under the assumption of comparison of similar documents (ceteris paribus). Finally, about mentions in patent applications, I have not found evidence of its association with social attention in Twitter when compared with similar documents.

## 5. Discussion

This work tries to explain the social media attention of a scientific research through some other social mentions and bibliometric characteristics of the paper. For this, publications in disciplinary journals of Clinical Medicine were used. Twitter is a social media platform with the potential to help scientists disseminate health-related research for policy impact (Kapp et al., 2015). In this field it is common for some research to be mentioned in public policy reports. In this respect, I obtained significant evidence that the typology of paper with potential application in public policies is a minor factor that contributes to the social media attention of a research (Twitter mentions). This typology is characterized by a high immediacy application to social problems, a rapid incorporation to knowledge and a rapid aging (Rowlands, 2009).

Among the characteristics of the research, I have included another typology of the paper, distinguishing between research article and review. A review paper is a highly valuable type of research output because it puts research into context for news reporters, policymakers, the public, and other researchers (Grant and Booth, 2009). However, in the analyzed dataset, I have found evidence of its weak association with social attention in Twitter when compared with similar documents.

There is significant evidence that the factor that contributes the most to the social media attention of a research (Twitter mentions) is the number of mainstream news. Thus, an additional mainstream news mention is associated with 0.67 increase in the number of mentions in Twitter, everything else equal. Moreover, an additional mention in Wikipedia is associated with 0.56 increase in Twitter mentions, while a reference in public policy documents is associated with 0.34 increase in the number of mentions in Twitter, ceteris paribus.

Interestingly, the average number of Twitter mentions for COVID-19 papers is 2.15 times higher than for not COVID-19 papers (more than double), everything else equal. However, in relative terms, the COVID-19 topic contributes to social attention on Twitter a 44% in relation to the number of news when comparing documents with similar characteristics.

Furthermore, the average number of Twitter mentions for a review paper is 1.25 times higher than for a research paper (25% higher), while for a paper mentioned by an expert is 1.72 times higher than for those not mentioned (72% higher for each recommendation), everything else equal.

About mentions in patent applications, I have not found evidence of its association with social attention in Twitter when compared with similar documents. A possible explanation for this result could be the following. This variable is more related to academic citations than to social media attention. That is, the innovation and business transfer, although being very relevant, It is difficult to communicate through social media. In this way, the most cited research is not necessarily the one that receives the most social attention (Dorta-González and Dorta-González, 2022).

Thus, citation and social attention do not correlate with each other and, therefore, they measure different dimensions in the impact of research results. This is something that has been pointed out in the literature (Bornmann, 2015; Costas et al., 2015) and means that altmetrics might capture diverse forms of impact which are different from citation impact (Wouters et al., 2019)

As a final consideration, altmetrics data (Altmetric, 2020) have the advantage of measuring different types of impacts beyond academic citations. They also have the potential to capture earlier impact evidence. This is useful in self-evaluations. Nevertheless, social attention of research must be used cautiously because it could provide a partial and biased view of all types of societal impact. For this reason, it should be avoided when evaluating researchers, especially in recruitment processes and internal promotions. In this work, social mentions were used to study the phenomenon of social attention of research itself and not to evaluate the researchers.

**Funding and/or Conflicts of interests/Competing interests**

Funding: This research received no external funding.

Conflicts of Interest: The author declares no conflict of interest.